# Quantum Machine Learning for Credit Scoring

**Nikolaos Schetakis [1,2,*], Davit Aghamalyan [3,4], Michael Boguslavsky [5], Agnieszka Rees [5], Marc Raktomalala [6], Paul Griffin [3]**

[1] Alma-Sistemi Srl, 00012 Guidonia, Italy
[2] Quantum Innovation Pc, Chania, 73100, Greece
[3] School of Computing and Information Systems, Singapore Management University, 81 Victoria street, 188065 Singapore
[4] Agency for Science, Technology and Research (A*STAR), Materials Science and Chemistry, Institute of High Performance Computing, 1 Fusionpolis Way, 16-16 Connecis, 138632 Singapore
[5] Tradeteq Ltd, London, UK
[6] Sim Kee Boon Institute for Financial Economics, Singapore Management University, 50 Stamford Road, 178899 Singapore

* Correspondence: nsx@alma-sistemi.com;

**Abstract:** In this paper we explore the use of quantum machine learning (QML) applied to credit scoring for small and medium-sized enterprises (SME). A quantum/classical hybrid approach has been used with several models, activation functions, epochs and other parameters. Results are shown from the best model, using two quantum classifiers and a classical neural network, applied to data for companies in Singapore. We observe significantly more efficient training for the quantum models over the classical models with the quantum model trained for 350 epochs compared to 3500 epochs for comparable prediction performance. Surprisingly, a degradation in the accuracy was observed as the number of qubits was increased beyond 12 qubits and also with the addition of extra classifier blocks in the quantum model. Practical issues for executing on simulators and real quantum computers are also explored. Overall, we see great promise in this first in-depth exploration of the use of hybrid QML in credit scoring.

## 1. Introduction

The quantum revolution, empowered with quantum computers, holds a promise to disrupt several scientific areas and industrial sectors including the financial sector [1-4]. Quantum computers are envisioned to bring exponential speedups compared to their classical counterparts [5]. An exciting quest for finding a practical problem where such computational advantage can be obtained has been ongoing for last 2 decades and is considered to be the "holy grail" of quantum technologies. After demonstrating such possibility on very specific problems [6,7] with currently available noisy-intermediate scale devices (NISQ) [8,9], the community came to a realization that it lacks having a class of practically useful, real-world problems where practical advantage could be obtained. In terms of quantum computation, we are currently in the so-called NISQ era [8,9], where the number of available noisy qubits is ranging from ten to the hundreds. It is crucial to realise, that scaling the number of qubits is a key challenge as it will equip quantum computers with fault tolerance, where the elementary information unit, a logical qubit*, is robust to noise [1]. That said, modern computational finance [10-12] offers a rich plethora of computational problems with which one can explore the application of quantum computers for obtaining the "holy grail". The idea of applying quantum physics to finance has been shown to be a fruitful one, with such fundamental examples as American option pricing [13] or even the ubiquitous task of portfolio optimization [14-15]. Furthermore, statistical mechanics can be used to study financial risk, where different tools borrowed from the fluid dynamics for solving non-linear differential equations can be employed to analyze foreign exchange markets [16]. The field which greatly benefits from applying concepts of quantum physics, quantum computing and Quantum Machine Learning (QML) to the problems of computational finance has been coined as "quantum finance" [2-5]. Important problems of computational finance include the risk management and the pricing of exotic derivatives using Monte-Carlo models, modelling financial markets with differential equations, credit scoring, portfolio optimization, understanding market trends with machine learning methods [17-19].

Since in this paper we are going to focus mainly on the task of credit scoring in what follows we briefly summarize why credit scoring [20,21] is important and what are the implications in succeeding in such a task. This paper is focused on the financial task of credit scoring [20,21]. Credit expands the economy by allowing firms to borrow capital at reasonable rates; however, market frictions such as information asymmetries may prevent capital from flowing to firms with profitable projects. Creditworthiness of large potential borrowers is typically assessed with public credit ratings; on the other hand, small companies are usually not covered by rating agencies and may suffer from reduced availability of credit. Over time financial institutions have gone from using "soft" qualitative company information to developing advanced techniques to quantify and manage credit risk across different product lines, especially with the growth of derivative instruments with intrinsic default risk exposure.

Developing and improving sophisticated Machine Learning (ML) credit scoring models, especially for small and medium-sized enterprises (SME), is a vibrant area. Currently even the best credit scoring models only reject 90% of companies that become insolvent (type II error) while rejecting 15% of companies that remain healthy (type I error) [Source: Tradeteq Ltd]. Even a small improvement in credit scoring model efficiency can translate to significantly lower rejections for good SME and lower risk for those funding the loans. ML [22,23] aims to classify, cluster and recognise patterns in datasets. Fundamental improvement to ML or QML could lead to significant economic benefits, and the community has been putting a big effort into applying different ML and QML frameworks to the financial sector with focus on real world financial applications and datasets. The scientific discipline which aims at understanding fundamental limits of data analysis and learning, imposed by the fundamental laws of quantum physics, is called QML [24-26]. QML is usually considered for tackling four different tasks, which can be defined through the data and algorithm being classical or quantum as follows: quantum data and quantum algorithm, classical data and quantum algorithm, quantum data and classical algorithm and classical data quantum algorithm. QML holds the promise to deliver practical advantage for data analysis. Such expectation has recently grown even higher for financial datasets.

In order to push the frontiers of obtaining the "holy grail" with QML in finance one has to keep open mind and open the stage for algorithms which are heuristic in nature. Here, by heuristic, we mean an algorithm which has no proven theoretical guarantees. At the same time, it is well understood that heuristic algorithms perform well on certain classes of problems where, through extensive collaboration between the experts from different areas, a deep domain knowledge is extracted. In this paper, despite our algorithm being inherently heuristic we provide an extensive benchmarking for our results by comparing the performance of our FULL HYBRID QNNs with that of the best-known classical ML NNs such as linear regression, logistic regression, vector machines and random forest models. Inspired by the idea that QML methods have a huge potential of obtaining practical advantage, we focus in this paper on models previously developed by authors, the FULL HYBRID classical-quantum neural networks architecture [27], to perform supervised learning on the problem of credit scoring in finance [20,21]. Large companies such as Microsoft or Citibank have many analysts providing credit ratings, but SME are not often analysed by rating agencies, and credit scoring is only derived from company and market data, if any. The first quantitative calculations of credit scores were by Altman in the late 1960's; his Z-score model estimates a linear combination of financial ratios and uses the statistical method of discriminant analysis to predict publicly traded company defaults within two years, in the manufacturing sector [20]. More recently, machine learning approaches allow for automated credit scoring for a broader coverage of attributes of small companies that combine company, accounting, and socio-economic information. Improving the machine learning algorithms is thus an important element to providing credit risk transparency. Optimal feature selection for credit scoring datasets has been suggested in Ref. [28]. This approach is based on an unconstrained binary optimization (QUBO) model. Comparison with such well-established methods as recursive feature elimination showed that QUBO feature selection resulted in smaller feature subset with no loss of accuracy. It is interesting to note that some researchers have benefited from thinking of quantum inspired algorithms. For instance, In Ref. [29] a quantum-inspired Neuro Evolutionary Algorithm with binary-real representation has been studied. The method performed equally well compared to the other classical ML methods. In this paper we apply our previously developed tools [27] to the problem of credit scoring. In our previous work [27], by combining such approaches as hybrid-neural networks, parametric circuits [32,33], and data re-uploading [30,31], we create QML inspired architectures and utilise them for the classification of non-convex 2 and 3-dimensional figures. An extensive benchmarking of our new FULL HYBRID classifiers against existing quantum and classical classifier models reveals that our novel models exhibit better learning characteristics to asymmetrical Gaussian noise in the dataset compared to known quantum classifiers, and perform equally well for existing classical classifiers, with a slight improvement over classical results in the region of the high noise. We highlight that general performance on synthetic datasets tends to be better compared with real world datasets as, in synthetic datasets, one can benefit, for instance, from the spherical symmetry of the problem. The paper is organized as follows. In Section 2, we explain the theory behind the classical models used for credit scoring, describe the quantum models that have been previously used and developed by authors and make some remarks on faithful benchmarking. In Section 3, we describe the dataset, the experimental setup and the results. Section 4 is devoted to our experiments which were run using the Pennylane emulator, here we give more details on the structure of QNN, outlining hyperparameters, number of epochs, learning rate and so on. We present results in Section 5, where we also summarise a very detailed benchmarking with best classically performing ML models, and also share data on runtime for our FULL HYBRID architecture. The final Section is devoted to summary and future directions.

## 2. Materials and Methods

In this section we review current classical and quantum models used for credit scoring and set the groundwork for our own novel hybrid classical/quantum model. Firstly, to assess and benchmark the models in credit scoring, the crucial trade-off is between Type I (false positive) and Type II (false negative) errors. This trade-off can be summarized by a ROC (Receiver Operating Characteristics) curve (Fig. 1). Where a false positive is rejecting a company for a loan that it would pay back, and true positive is rejecting a loan where the company would have defaulted. Ideally, a perfect model would not reject a good company (0.0 on the x axis on the graph in fig. 1) and would reject all companies that would default (1.0 on the y axis on the graph in fig. 1).

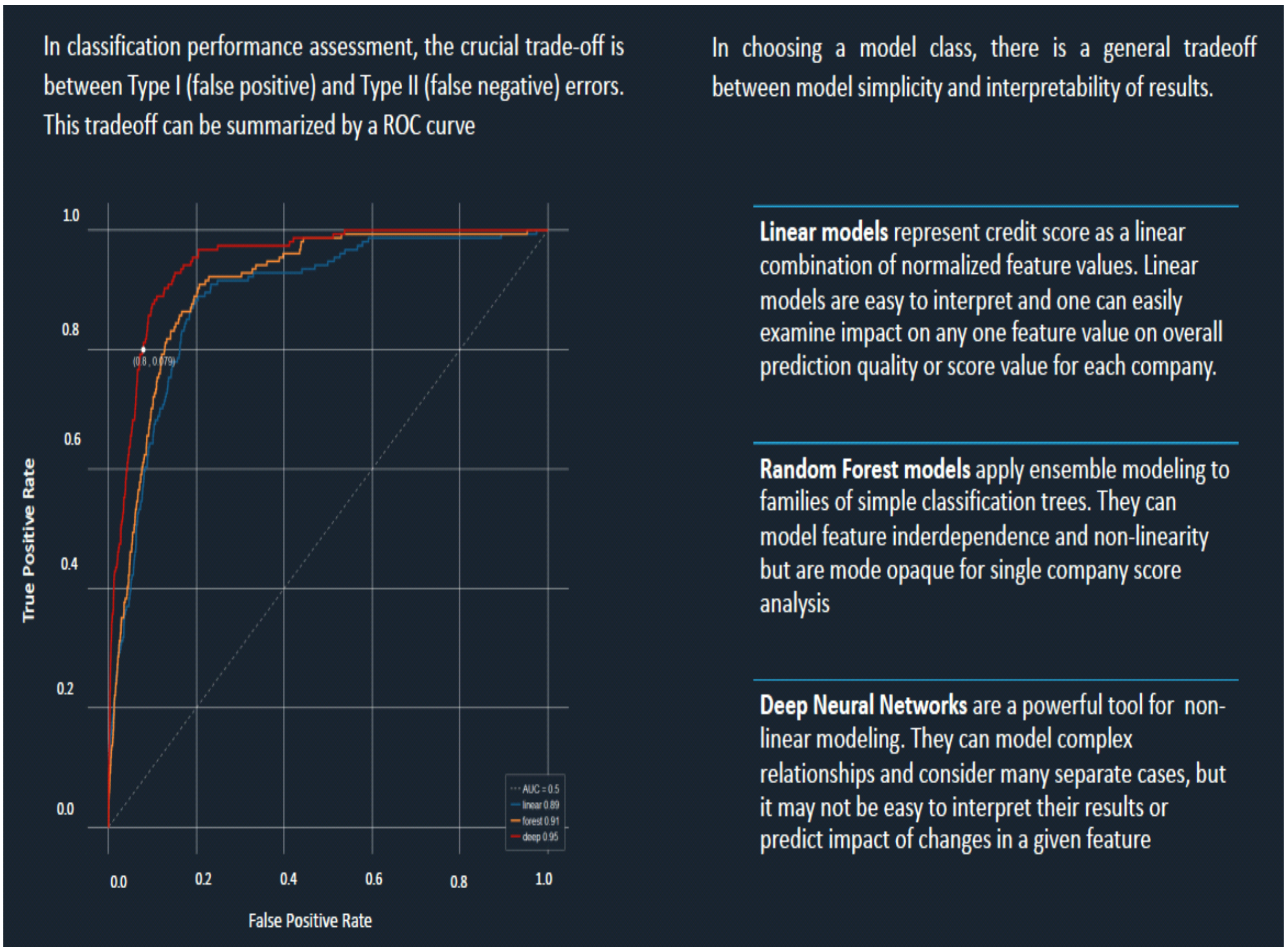

**Figure 1.** The ROC curves of three models used for credit scoring. (Source: Tradeteq Ltd.)

The area under the ROC curve (AUC) is calculated by summing the area under each ROC curve in Figure 1 and normalising the areas to 1.0. AUC varies between 0, classifier always giving the wrong answer, and 1, perfect classifier performance, with 0.5 for a uninformative classifier. This is a useful criterion to compare models and is the performance metric used throughout the paper.

### *2.1. Classical models*

Here we briefly outline classical ML attempts towards credit scoring where we focus on the methods used and on the accuracy of prediction of a company defaulting on a loan. As a machine learning problem, credit scoring is typically formulated as a binary classification problem with highly imbalanced data (most companies do not default). The data used has a number of features for a set of companies observed at a specific time, T0, and an outcome observed at a later time, T1. The outcome is usually treated as binary highlighting, class “1” if a company had some kind of adverse event between T0 and T1, such as companies in bankruptcy proceedings or in administration, and class “0” for all other companies in the dataset. As most companies do not suffer adverse events in most periods, the data is strongly imbalanced to class “0”. A review of all the studies on machine learning for credit scoring is a subject large enough for a standalone paper. Indeed, we refer the reader to systematic academic reviews [39]. The first models used relatively small company sets (hundreds of companies) and just a few manually constructed accounting features; thus linear techniques such as linear discriminant analysis and support vector machines were used. Later, researchers obtained larger datasets with more companies and features, and applied linear regressions, decision trees, fuzzy logic, ensemble models, and neural networks to the problem. With large enough datasets, modern ensemble techniques, such as boosted trees, and neural networks perform broadly on par. However, boosted trees are often preferred in practice due to their better explainability and stability. Choosing a classification model for credit scoring is a challenging task, and conflicts often arise when comparing performance. For example, linear discriminant models for predicting bad loans are found to perform better than neural networks for some data and opposite for other. These different outcomes are difficult to assess but possible explanations include differences in sample sizes, transformation functions applied to the data, model parameters, or network topology. Also, the chosen performance metric matters; traditional statistical methods seem to perform as well as neural networks if one considers the total percentage of correct identification but, if identifying bad loans is the main goal, then neural networks have been found to perform better [35]. The search for an effective "generic universal" model may be elusive but failure to incorporate situational data and consider local economic circumstances affects the ability to develop relevant powerful predictive models. In that regard, in this paper, we consider the best classical Singapore country model for classification of two groups of potential SME borrowers (called "best classical model" in the rest of the paper). The best classical model is described in the experiment section below. The performance metric throughout the paper is the AUC (see benchmarking section below) and this model achieves a score of 0.73 (see results section). XGBoost has been found to be the best model for this dataset but, as we want to compare neural networks, this model is not used in this paper.

### *2.2. Quantum models*

In the book chapter co-authored by M. Boguslavsky, P. Griffin et al. [34], the authors introduce a new framework for addressing business problems with quantum computing, assessing classes of problems which could benefit and show a use case for quantum machine learning (QML) algorithms. The authors outline two frameworks for quantum neural networks: (i) a 2-qubit perceptron inspired by the Entropica Labs algorithm for classification of cancerous cells and (ii) a hybrid neural networks where it is suggested to establish an interface between classical and quantum neural networks using PYTORCH and Qiskit [36]. In finance, there are extensive overviews/reviews for quantum computing and QML applied to finance [2-5]. In all these overviews, credit scoring is mentioned as a problem which the current community is targeting to solve by making use of QML algorithms.

Now we would like to describe our own novel architecture used for the experiments in this paper. It is a FULL HYBRID (FH) quantum neural network model consisting of three different approaches based on: hybrid-neural networks [36], variational circuits (VC) [32,33] and data-reuploading classifier (DRC) [30,31]. First, classical data has to be encoded into quantum states. Angle embedding is commonly used to load the data [x1, x2] into a qubit. Starting from an initial state vector, typically $|0\rangle$, a unitary operation U(x1, x2, 0) is applied and a new quantum state is formed which can be described by a new point on a Bloch sphere. Padding with 0 is required when dealing with 2 or more dimensional data, for example, loading higher dimensional data [x1, x2, x3, x4, x5, x6] can be broken down into sets of three parameters: U(x1, x2, x3), U(x4, x5, x6). We use a Rx gate for angle embedding in our experiments. Hybrid neural network classical-quantum classifiers are formed by connecting a number of classical and quantum neural networks in series. This architecture takes advantage of the specific capabilities of both types of neural networks and also benefits from being able to have the number of features in the initial classical layers exceed the number of qubits in the quantum layer, instead of being limited to one qubit for each feature. To create our hybrid classical-quantum neural network a hidden layer is implemented utilising a variational quantum circuit (Fig. 2). A variational classifier (VC) is a quantum circuit (also called a "parameterized" circuit) consisting of the data embedding layer followed by parameterised gates such as rotation gates and entangling layers (CNOT gates that entangle each qubit with its neighbour). The quantum properties such as the rotation angles for the quantum gates are trainable parameters. DRC is introduced by replicating the VC into more blocks (for example 2nd block in Fig. 2). To combine a VC circuit with the DRC technique we define a block (B) as a sequence of data embedding and entangling layers (L). By adding many blocks, we re-introduce the input data into the model. Our novel approach is to use the DRC technique combined with a VC in a single model in the quantum part of the hybrid classifier. This novel combination is expected to provide greater robustness to our results after making observations on 2d and 3d synthetic non-convex datasets [27]. The advantage brought by the VC approach for binary classification is in increased robustness against noise, however it struggles to capture the complicated patterns in the prediction grid diagrams. Conversely, the DRC approach has good abilities to capture complex grid structures but more sensitive to the noise in data. Consequently, it is expected that VC and DRC combined together will complement each other to get the best results. The power of the model lies in capturing the complex patterns in the data with robustness to data noise. We previously demonstrated [27] that, for synthetic datasets, FH architectures: (i) outperform several previously known quantum classifiers, (ii) perform equally well compared to classical counterparts and (iii) have an improvement over classical counterparts in regions of high noise in the dataset.

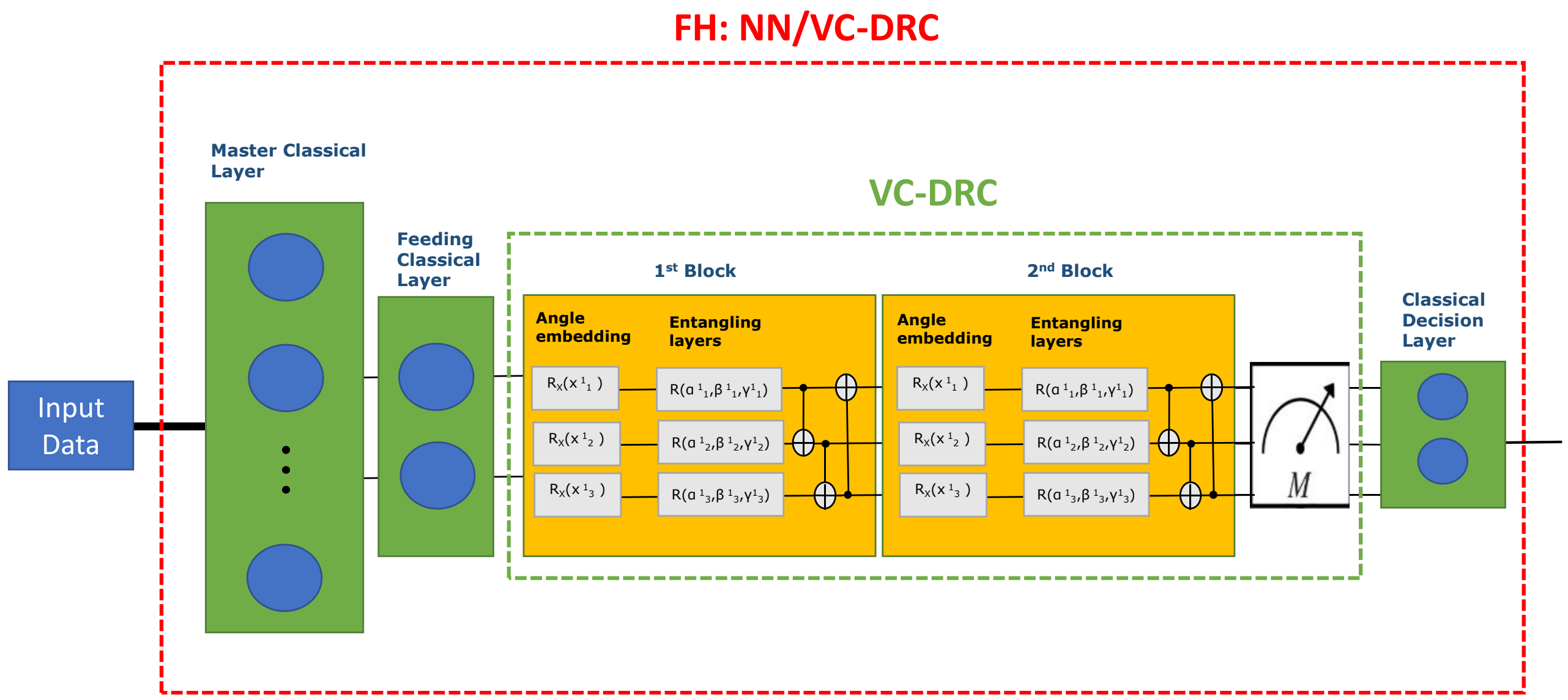

**Figure 2.** Block Diagram of the Full Hybrid (NN/VC-DRC) classifier where a VC-DRC circuit is placed after a classical neural network.

After computation on the quantum node is completed, measurement is performed. The measurement outcome is the expectation value of a Pauli observable for each qubit. The measurements are passed to the classical decision layer which makes the final prediction label of the binary classifier. Another architecture explored for this problem area is to append the master classical layer to the quantum layer instead of having it first (i.e. FH:VC-DRC/NN). However, as the results for the dataset under investigation in this paper were found to be better with the FH:NN/VC-DRC model, we only focus on this model and going forward call this the FH model.

### *2.3. Data description*

In this paper, we compare the performance of quantum machine learning classification algorithms with their classical counterparts, applied to a real-life credit scoring dataset. The dataset originates from various Singapore institutions — the Accounting & Corporate Regulatory Authority (ACRA), a statutory board under the Ministry of Finance of the Government of Singapore, the Singapore Land Authority, a statutory board under the Ministry of Law of the Government of Singapore, the Singapore Buildings database, Handshakes, a corporate data provider, and Tradeteq, a provider of data, technology, and software to the trade finance industry. The dataset covers nearly 2,300 SME firms incorporated over the 1940 – 2016 period, active and healthy in 2016, with 94% of the firms incorporated since 1990 and distributed mainly across seven industry sectors. The most important peculiarities of SME datasets are that firms are privately held and that there is limited information about the financial situation of borrowers — only accounting data is available, and no information from rating agencies nor financial markets prices. These data limitations restrict the modelling choices of an SME portfolio to binary default or no default models and panel data analysis, rather than time series analysis. For each small business, the panel of 24 primary features, includes the year of incorporation, accounting and operating information, geo-sociological data, and an indicator of whether the company defaulted or not. The firm is statutorily deemed “Healthy” (class 0) on October 1, 2016, and its status, in case of default (e.g. compulsory winding down, receivership, or under judicial management), is statutorily changed to “Unhealthy” (class 1) over the following two years, between October 1, 2016 and October 1, 2018. The dataset size is limited by the number of class 1 examples, i.e. the companies that defaulted. We sampled all 246 class 1 companies in the period and added to it a random subsample of 2000 class 0 companies. Therefore, the dataset is highly imbalanced. For the experiments, due to the limitations of the computing hardware, the maximum number of features used is 21.

### *2.4. Experiments*

In this section we discuss the experiments performed on the Singapore data over a period of two years of simulations trying out many different models, activation functions, epochs, and other parameters. Here we discuss the best quantum and classical counterpart (CC) models found, the hardware and software used, the hyperparameter configurations, model executions and the processing of the experiment outputs. We used an FH:NN/VC-DRC model (Fig. 2) where the first part is a classical neural network, followed by the VC-DRC circuit and a final decision layer, a single neuron layer with a sigmoid activation function. The first classical layer «Master Classical layer» has 21 neurons, equal to the number of features used in our dataset. The second classical layer «Feeding Classical Layer» has the same number of neurons as the number of qubits. Moreover, the classical neural network can contain an arbitrary number of layers, and each layer can contain an arbitrary number of neurons, but the last layer (Feeding classical layer) must have the same number of neurons as number of qubits. In our 2D case the classical NN, consists of a 2-neuron layer with a rectified linear activation function (ReLU) (Master classical layer), followed by a 2-neuron layer with a Leaky ReLU activation function (Feeding classical layer). Finally, a classical decision layer «Decision Classical layer» makes the prediction. All layers have a ReLU activation function except for the final decision layer with a sigmoid activation function. After each classical layer, to avoid overfitting, we added a dropout layer to both quantum and CC models with 10% dropout rate. Before the data is ingested into the model, it is pre-processing using a standard pipeline along with some proprietary processing. This is the same processing that is used in the best XGBoost classical models.

Data encoding into quantum states used angle embedding (see 2.THEORY). As simulators were used throughout, there is no quantum computer noise, and no additional noise was introduced in these experiments. See the discussion section for considerations on how noise may affect the results on real quantum computers. Pennylane, an open-source software framework for differentiable programming of quantum computers, was used to build the models. The computer used for these experiments was a PC with 64GB of RAM and an AMD Ryzen 7 processor. We tested the quantum model performance versus the number of qubits and number of blocks and against the CC model. The training dataset has 1798 rows of data split into a validation set of 180 rows (10%) and a test set of 270 rows (15%). Throughout the simulations, the same training, validation, and testing dataset is used. For every configuration, we used the average outcome of ROC/AUC score of 5 simulations. The hyperparameters used are summarised in table 1 and were kept consistent throughout the study except where explicitly mentioned.

**Table I.** Hyperparameters used in this study.

| | | |
|---|---|---|
| Epochs | 350 | Number of complete passes through the training dataset |
| Dropout rate | 0.1 | Probability of training a given node in a layer, 1.0 = no dropout, 0.0 = no outputs from the layer |
| Learning rate | 0.001 | Step size at each iteration while moving toward a minimum of a loss function |
| Optimizer | SGD | Stochastic Gradient Descent |
| Batch size | 16 | Number of training samples to work through before the model's internal parameters are updated |

For a fair comparison, the number of epochs for quantum and classical models were kept identical at 350. This number was chosen after observing that, on average, after 350 epochs, overfitting was observed. However, the classical model improved if training was increased up to 5,000 epochs. The number of DRC blocks was increased from 1 to 10 and the number of qubits was increased from 6 to 18. The batch size was reduced to 16 training examples per iteration due to memory restrictions. The optimizer we used is a Stochastic Gradient Descent (SGD) optimizer. The loss function is binary cross-entropy, and the optimising parameter is ROC/AUC as described in 2. THEORY. Throughout the simulations, we used the same training, validation, and testing dataset for consistency. For every configuration specified, we used the average outcome of the ROC/AUC score over 5 simulations. Our simulations were restricted to a maximum number of 18 qubits due to execution time constraints (see scaling results in 5. RESULTS and 6. DISCUSSION).

## 3. Results

In this section we discuss the experiments We now show the experimental results comparing the overall ROC/AUC of the FH model to the CC model and show the quantum model's behaviour as we scale up the number of qubits and the number of processing blocks.

The quantum model achieved an ROC/AUC of 0.75 (Fig. 3 black dots and Fig. 5 right column) whereas the CC model achieved an ROC/AUC of 0.73 (Fig. 3 orange line). However, letting the classical counterpart model train up to 5,000 epochs can lead to the highest score of 0.75 (Fig. 3 red line). This indicates that the FH model has the ability to achieve a higher score than its classical counterpart with fewer training epochs.

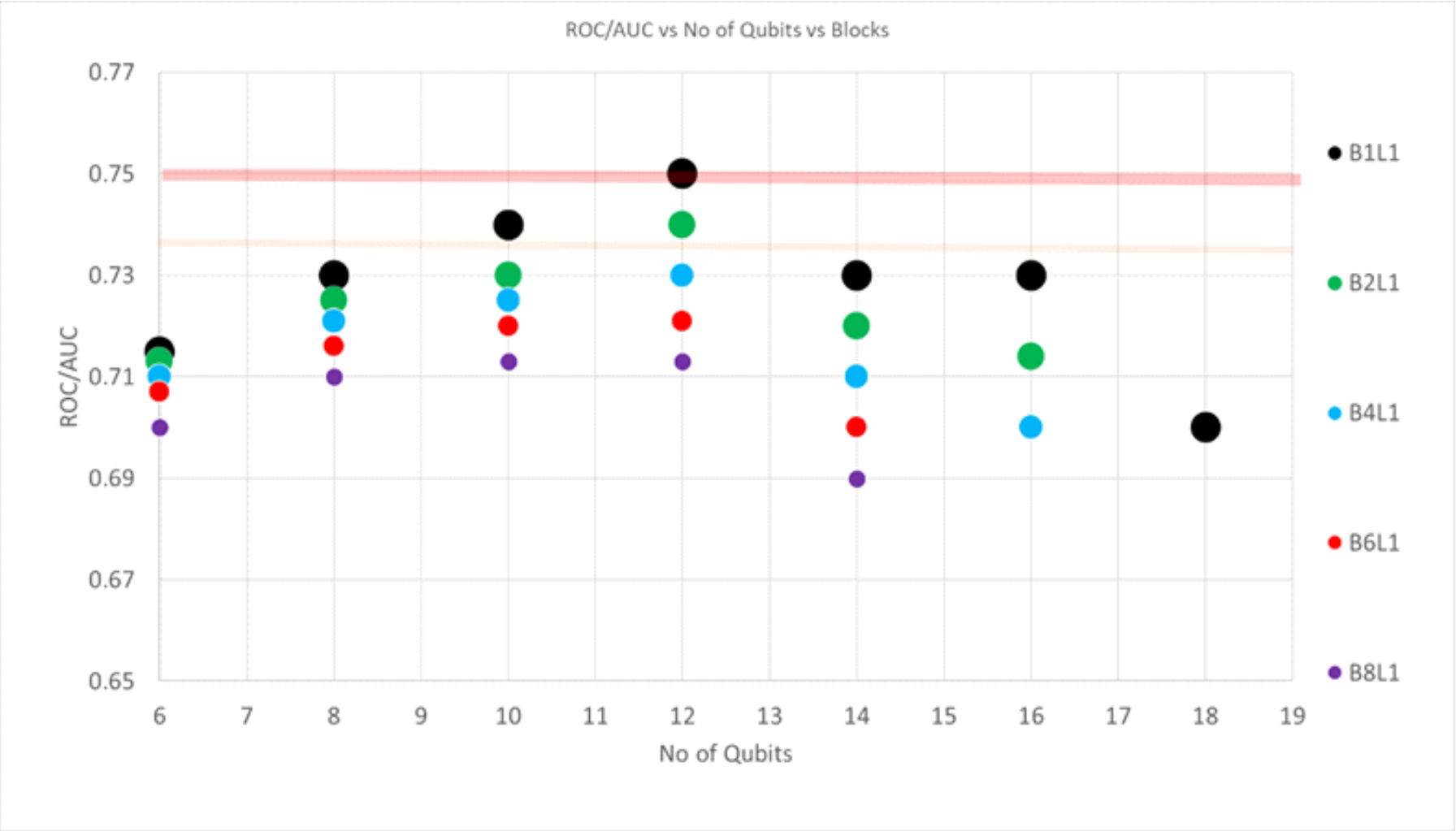

**Figure 3.** The FH model versus the number of qubits (x-axis) and number of blocks (coloured circles). The orange line denotes the best classical model ROC/AUC when trained for a maximum of 350 Epochs while the red line denotes the best classical model's ROC/AUC when trained for a maximum of 3,500 epochs.

Furthermore, in Figure 3, we see how the ROC/AUC of FH model performance changes with the number of qubits and number of blocks. For blocks=1 (black dots) we observe that ROC/AUC increases up to 12 qubits and then decreases. The same behaviour can also be observed when one increases the number of blocks with the highest results being achieved with no data-reupload. See the discussion section below for possible explanations.

The training processes that produce the highest ROC/AUC scores on the testing dataset are depicted in Figure 4 for both CC model (left and middle columns) and the FH model (right column). The top row depicts the final ROC/AUC, the middle and bottom rows respectively show the loss and ROC/AUC training process for the training and validation dataset. The FH model achieved the highest score with fewer training epochs for all simulations shown in Figure 4.

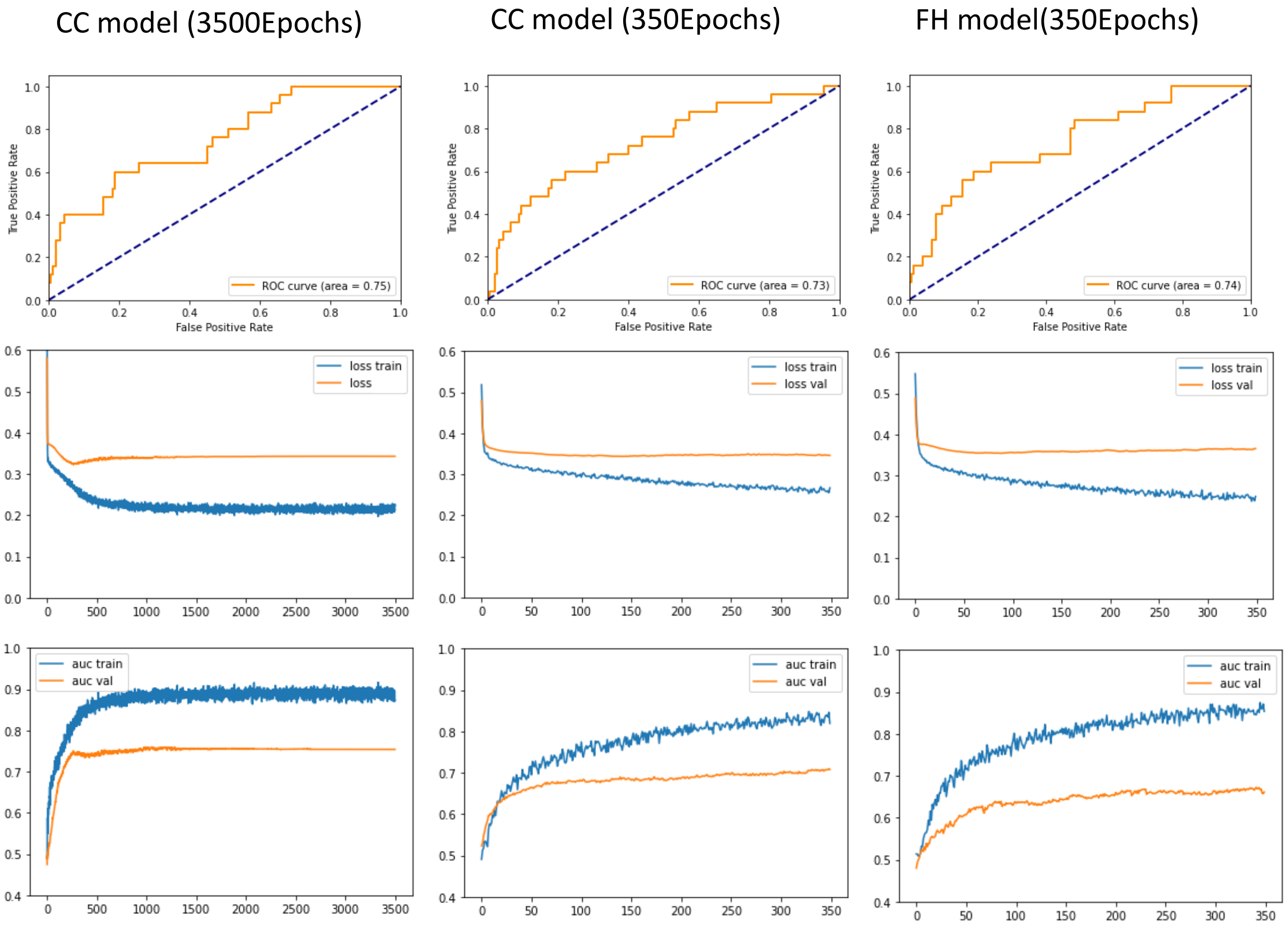

**Figure 4.** Results of both CC model (left and middle columns) and FH model (right column). Results for CC model are shown for different max number of training epochs: 3,500 epochs (left column) and 350 epochs (middle and right column).

Measuring execution times on the simulator allows estimating resource requirements for actual quantum computers. We observed that data embedding time doubles with each additional qubit, model execution time scales quadratically with the number of qubits and linearly with the number of blocks (Fig. 5). These observations are of importance for the use of real quantum computers and are discussed in the section below. We used up to 20 qubits for this test.

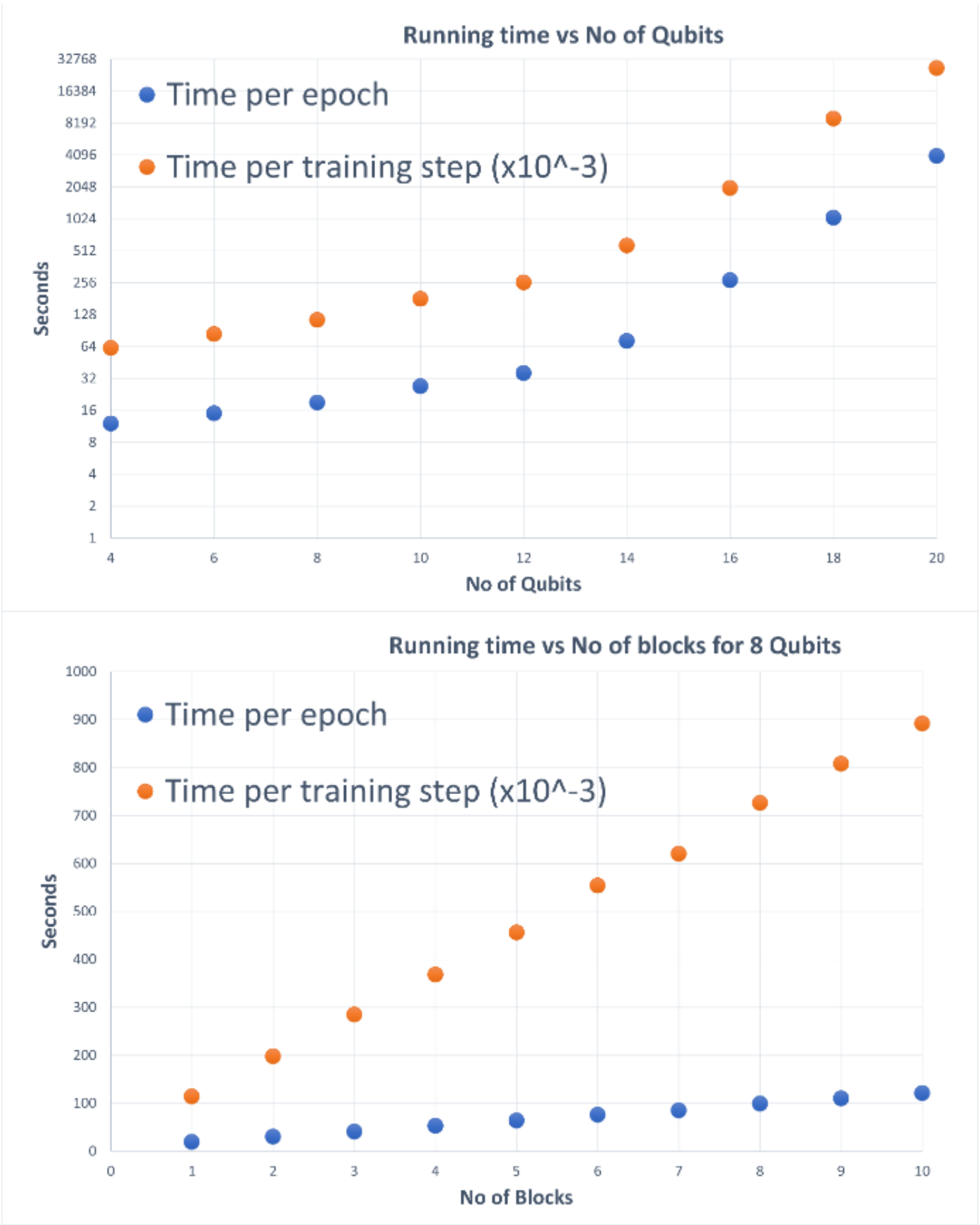

**Figure 5.** The execution time for the FH model on the simulator per epoch and per training step for an increasing number of total qubits (top), and for an increasing number of blocks with a constant number of 8 qubits (bottom).

## 4. Discussion

Firstly, we would like to highlight that what we propose is a novel approach using a new architecture combining typical quantum variational circuits combined with a data re-uploading technique in a hybrid (classical-quantum) neural network (our new architecture is named under common name FULL HYBRID (FH) [27]). We will now discuss the comparison of our FH and CC models as well as the execution times and limitations. We found the accuracy of the FH model to be equal to the CC model. Also, the FH model required significantly less epochs to train than the CC model by a factor if 2 or more (Fig. 5). Given that training time is a significant business driver this result is very interesting and grounds for optimism for practical quantum advantages in the future. That the results are satisfactory, even when using a classical feeding layer to reduce the number of features fed into the quantum layer (limited by the number of qubits), is another positive finding. Currently, the number of qubits is limited in the simulator by memory requirements growing exponentially with the number of qubits. For real quantum computers the physical number of qubits is also limited by the available hardware and extra qubits needed for error correction. Consequently, this is the most practical FH architecture we can use for this specific dataset. In the future, increased computing power will enable us to benchmark with more qubits and produce more results. The unexpected finding that increasing the number of qubits above 12 (Fig. 3) reduced the accuracy could be related to barren plateau issues but is to be investigated in future studies. The effect of adding more blocks, i.e. DRC, reducing the accuracy (Fig. 3) could be due to our dataset not having features that have a trigonometric (i.e. sine) structure thus re-entering the data to the model doesn't improve its performance [13]. It is not due to overfitting as the training is stopped before overfitting occurs. We note that finding the best learning gradients in a non-convex landscape (for the problem structure) is an open question even in the classical machine learning community and no resolution has been discovered so far. To this end, one possible solution is to use Quantum Convolutional Neural networks and, moreover, the use of a data re-uploading circuit could possibly overcome this problem based on the prima facie argument that, since data is introduced many times to the network, the solution is forced from any local minima. Another promising method is to characterize the landscape by computing the Hessian of the loss function where, since the eigenvalues of the Hessian loss function quantify directly the local curvature of the loss function, we can adjust the learning rate of our model for faster convergence during the training process. For simulators, we observe the exponential time complexity of increasing the number of qubits and blocks (Fig. 5), both of which would not be a problem for a real quantum computer as long as the circuit width (number of qubits) and circuit depth (number of gates) are within the specification of the quantum computer. However, another general issue for QML is the need to make many executions of the quantum circuit, for 350 epochs and a training set of 1798, this amounts to 629,300 circuit executions. This is not an issue for simulations run on a standalone computer. However, for a quantum computer accessed on a cloud platform, the overall time taken may not practical, for example it may take 7 days in total if each execution takes 1 second due to network overheads and queuing time on a shared quantum computer. Quantum computer providers such as IBM have very recently introduced mechanisms such as Qiskit runtime that enables the execution of the classical and quantum code to be run as one unit reducing the overall execution time by 120 times, potentially bringing 7 days down to 1.5 hours. The dataset size is limited by the number of class 1 examples available. This has an impact on the most appropriate model but does not lead to overfitting. A much larger dataset such as the UK's with 1000+ defaults each year and over 4 million companies would be interesting to use with more complex models and finer grained risk periods. We also note that hyperparameters for the FH and CC models could potentially be tuned further.

## 5. Conclusions

In conclusion, the use of hybrid quantum/classical models is promising given the ease of obtaining comparable results to a purely classical counterpart and with much fewer epochs for training. We have investigated the practical issues of using the models on simulators and on real quantum computers and expect that, with even modest improvements in hardware expected of over 4,000 qubits by 2025 [37], along with software improvements such as runtime environments [38], real advantages, at least in model training, will be achieved. It is also possible that improvements in accuracy may also be observed due to the resilience in the FH model. However, the best classical model for Singapore data is still XGBoost and this work can be extended to other datasets, with more features using more qubits. Furthermore, this study shows that anyone in the machine learning community can relatively easily experiment with QML for their own problems. The next step of moving to real quantum hardware may also prove interesting with the introduction of quantum noise possibly removing the need for dropout layers. The future for QML is very exciting. All the codes used in the manuscript will be provided under a reasonable request.

**Author Contributions:** For research articles with several authors, a short paragraph specifying their individual contributions must be provided. The following statements should be used "Conceptualization, X.X. and Y.Y.; methodology, X.X.; software, X.X.; validation, X.X., Y.Y. and Z.Z.; formal analysis, X.X.; investigation, X.X.; resources, X.X.; data curation, X.X.; writing—original draft preparation, X.X.; writing—review and editing, X.X.; visualization, X.X.; supervision, X.X.; project administration, X.X.; funding acquisition, Y.Y. All authors have read and agreed to the published version of the manuscript." Please turn to the CRediT taxonomy for the term explanation. Authorship must be limited to those who have contributed substantially to the work reported.

Conceptualization, P.G.; methodology, M.B.,A.R. and D.A. , N.S. ; software, D.A.,N.S. and A.R.; validation, M.B.,A.R. and M.R., formal analysis, P.G.,M.R.; investigation, M.R..; resources, N.S. and D.A.; data curation, A.R.; writing—original draft preparation, D.A..; writing—review and editing, P.G., M.R. and D.A.; visualization, N.S.; supervision, P.G.; project administration, P.G.; funding acquisition, P.G. and N.S. All authors have read and agreed to the published version of the manuscript.

**Funding:** N.S. would like to acknowledge funding support from the European Union's Horizon 2020 research and innovation programme EΥE under the Marie Skłodowska-Curie grant agreement No.101007638. The research project was supported by the Artificial Intelligence and Data Analytics (AIDA) scheme, which provides funding support to strengthen the AIDA ecosystem in the Singapore financial sector. The AIDA scheme is part of the Financial Sector Technology and Innovation (FSTI) scheme under the Financial Sector Development Fund administered by the Monetary Authority of Singapore (MAS). Any opinions, fundings and conclusions or recommendations expressed in this material are those of the author(s) and do not reflect the views of MAS. D. A. would like to acknowledge the funding support from Agency for Science, Technology and Research (#21709)

**Acknowledgments:** We would like to thank Kishor Bharthi for supporting our work from its early stages.

**Conflicts of Interest:** The authors declare no conflict of interest.